\documentclass[aps,prd,onecolumn,groupedaddress,preprintnumbers,superscriptaddress,noshowpacs,nofootinbib]{revtex4}
\usepackage{amsmath,amssymb,latexsym}
\usepackage[dvipdfmx]{graphicx}
\usepackage{color}
\usepackage{mathrsfs}

\usepackage{enumerate}

\begin{document}

\title{
The essence of the Blandford--Znajek process
}

\author{Shunichiro Kinoshita}
\email{kinoshita@phys.chuo-u.ac.jp}
\affiliation{
Department of Physics, Chuo University, 
Kasuga, Bunkyo-ku, Tokyo 112-8551, Japan
}
\author{Takahisa Igata}
\email{igata@rikkyo.ac.jp}
\affiliation{Department of Physics, Rikkyo University, Toshima, Tokyo 175-8501, Japan
}

\preprint{RUP-17-21}

\date{\today}

\begin{abstract}
 From a spacetime perspective, the dynamics of magnetic field lines of
 force-free electromagnetic fields can be rewritten into a quite similar
 form for the dynamics of strings, i.e., 
 dynamics of ``field sheets''.
 Using this formalism, we explicitly show that the field sheets of
 stationary and axisymmetric force-free electromagnetic fields have
 identical intrinsic properties to the world sheets of rigidly rotating
 Nambu--Goto strings. 
 Thus, we conclude that the Blandford--Znajek process is
 kinematically identical to an energy-extraction mechanism by the
 Nambu--Goto string with an effective magnetic tension.
\end{abstract}

\maketitle

 \section{Introduction and summary}

 The rotational energy of rotating black holes is a promising energy source
 for the formation of relativistic jets, which are ubiquitous in astrophysics.
 The Blandford--Znajek process~\cite{Blandford:1977ds}, which is an
 energy-extraction mechanism by force-free electromagnetic fields, can
 efficiently achieve powerful energy fluxes and thus has been widely believed
 to be a viable mechanism to extract the rotational energy of a black hole.
 There have been a number of analytical and numerical investigations from
 various aspects during the four decades since this process was proposed.
 A detailed analysis of the extraction mechanism was in terms of the
 membrane paradigm~\cite{MacDonald:1982zz}.
 In recent years interpretations and explanations of this mechanism
 have been discussed 
 (see, e.g., Refs.~\cite{Komissarov:2008yh,Koide:2014xpa,Toma:2016jmz}). 
 Moreover, various numerical simulations have been developed and
 demonstrated~\cite{McKinney:2004ka,McKinney:2012vh,Penna:2013rga}.
 In this paper we will analytically reveal the essence of the energy-extraction
 mechanism in the Blandford--Znajek process from an alternative
 perspective.

 Recently, it has been elucidated that 
 rigidly rotating Nambu--Goto strings~\cite{Frolov:1996xw} twining around a rotating black hole can highly
 efficiently extract the rotational energy from the black hole~\cite{Kinoshita:2016lqd}.
 The order of the possible energy flux, namely the energy-extraction rate, 
 is given by the so-called Dyson luminosity%
 \footnote{For historical details, see Ref.~\cite{Barrow:2017atq}.}
 $\sim c^5/G_\mathrm{N}\simeq 10^{59}\mathrm{erg/s}$~\cite{Dyson:1963},
 consisting of only fundamental constants: the speed of light $c$ and
 Newton's constant $G_\mathrm{N}$, and
 a dimensionless string tension $G_\mathrm{N}\mu/c^2$ as a coefficient.
 Such energy flux and angular-momentum flux are locally determined by the locus
 where the string intersects the light surface associated with its
 angular velocity, at which the velocity of the corotating frame
 coincides with the speed of light.
 Moreover, a necessary condition for the energy extraction to occur is that the
 light surface enters into the ergoregion and the angular velocity of
 the string is less than that of the black hole.
 
 In this mechanism, if we replace the tension of the Nambu--Goto string with
 a typical magnetic tension of electromagnetic fields surrounding a rotating
 black hole $\sim B^2r_\mathrm{h}^2$, we can reproduce 
 the energy-extraction rate of the
 Blandford--Znajek process while assuming conventional
 values of the magnetic field and the black hole mass.
 This fact suggests that both the energy-extraction mechanisms are closely
 related and the magnetic field lines with magnetic tension play an
 essential role in the  Blandford--Znajek process.  
 The purpose of this paper is to exhibit 
 that such observations are exactly true in a
 quantitative and theoretical 
 sense as well as a qualitative and intuitive sense.
 We show that the energy-extraction mechanism by
 stationary, axisymmetric force-free electromagnetic fields is
 essentially identical to that by rigidly rotating Nambu--Goto strings.
 It follows that 
 {\em the Blandford--Znajek process, i.e., the
 energy-extraction mechanism from rotating black holes via force-free
 electromagnetic fields, is the Penrose process for magnetic field
 lines with angular-momentum and energy transport mediated by their
 magnetic tension.
 }

 For this purpose, we will first reformulate the dynamics of force-free electromagnetic fields in
 accordance with a ``field-sheet'' formalism~\cite{Carter:1979,Uchida:1997a,Uchida:1997b,Gralla:2014yja}.
 In this formalism, magnetic field lines are fundamental objects to
 describe dynamics rather than electric and magnetic fields. 
 A time evolution of a magnetic field line can be regarded as a
 2D extended object in a spacetime.
 We will call such objects ``field sheets'', a name that was
 adopted in Ref.~\cite{Gralla:2014yja}, with a similar connotation to
 the term ``world sheet'' of strings in a spacetime.
 In the case of magnetically dominated force-free
 electromagnetic fields ($F_{\mu\nu}F^{\mu\nu}>0$) in particular,
 the field sheets become 2D
 timelike surfaces characterized by the electromagnetic field strength
 $F_{\mu\nu}$.
 We can recast the equations of motion for force-free
 electromagnetic fields, equivalent to the Maxwell equations with the
 force-free condition, in the
 equations of motion in terms of the field sheets.
 Thus, it turns out that the dynamics of field sheets for force-free
 electromagnetic fields is similar to the dynamics of a world sheet for a
 string, and they both belong to the same category of dynamics of
 a 2D surface in a spacetime 
 (a similar approach to magnetohydrodynamics and its phenomenological
 applications were discussed in, e.g.,
 Refs.~\cite{Christensson:1999tp,Semenov:2000vb,Semenov:2004ib}).
 This provides an insight into a correspondence between extraction
 mechanisms by force-free electromagnetic fields and rigidly rotating strings.

 Needless to say, electric and magnetic fields are not components of
 vector fields, but some of the components of a tensor field, i.e.,
 the electromagnetic field strength $F_{\mu\nu}$.
 Therefore, depending on 
 coordinate systems or reference frames, 
 their physical interpretations can
 change as well as their values.
 One may, for instance, move to a frame at which only magnetic fields can be
 observed even if there exists a net energy flow.
 Thus, the Poynting flux is only an interpretation of the energy flow in
 another frame.
 Furthermore, in a strong gravitational field, namely, in a curved spacetime, 
 the spacetime metric
 affects conventions or interpretations of the electric and magnetic
 field without physical significance.
 This means that explanations based on electric and magnetic fields will
 vary depending on the choice of frames.
 This seems to interfere with 
 the concise understanding of the mechanism.
 What we should emphasize is that the field sheet is a geometrical object
 irrelevant to any coordinate system as well as a physically
 intelligible object, thought of as the time evolution of a magnetic
 field line.
 It is expected that we can grasp the essence without
 suffering from coordinate systems.
 
 With the above perspective in mind, in Sect.~\ref{sec:3}, we focus on
 stationary and axisymmetric force-free 
 electromagnetic fields in a stationary and axisymmetric spacetime to
 examine the energy-extraction process from a rotating black hole.
 Since the basic properties of such electromagnetic fields have been
 widely examined and are well known in the literature 
 (see, e.g., Ref.~\cite{Frolov:1998wf} and
 references therein), we translate those
 into expressions based on the field sheet.
 We demonstrate that the field sheet of the electromagnetic field and the world
 sheet of the Nambu--Goto string have identical intrinsic properties such
 as their induced geometries. 
 In particular, the specific angular-momentum and energy fluxes per unit
 tension, flowing on each field sheet, are determined by local
 configurations of the magnetic field lines in the same manner as the
 string configurations.
 Most importantly, on the light surface these specific quantities depend
 only on its locus without global configurations determined by the
 equations of motion and must satisfy there identical
 relations regardless of whether magnetic field lines or strings;
 if the angular velocity of a magnetic field line is less than that of a
 black hole, the angular momentum can be extracted, and in addition if the locus
 where the magnetic field line crosses the light surface enters into the
 ergoregion, the energy can be extracted.
 (The importance of the light surface and other characteristic surfaces
 such as the Alfv\'en surface has been discussed in the literature~\cite{Takahashi:1990bv,Contopoulos:2012py,Thoelecke:2017jrz}.) 
 However, because the dynamics of magnetic field lines and strings are quite similar
 but different, both configurations cannot be identical in general. 
 This fact indicates that global configurations of the magnetic fields
 have little significance for this energy-extraction mechanism and a
 local, kinematical process in the ergoregion should govern the
 energy-extraction mechanism. 
 Hence, we can conclude that the essence of both the energy-extraction
 mechanism by the force-free electromagnetic fields and the rigidly
 rotating strings is identical.

 In contrast with the tension of Nambu--Goto strings, the magnetic
 tension can vary and should be determined by solving the equations of motion.
 This means that global configurations of the magnetic fields can affect
 the value of the magnetic tension.
 However, a role of the magnetic tension proportionally provides efficiency of
 extraction rate.
 Roughly speaking, the larger the magnetic tension is, the more
 efficient the extraction rate becomes. 
 In this process, the magnetic tension transports the angular-momentum
 and energy fluxes on the magnetic lines.

 The fact that the essence of the extraction mechanism is local
 kinematics in the ergoregion 
 irrelevant to global configurations offers some instructive
 insights to clarify the whole picture of the Blandford--Znajek process.
 It is not so significant whether magnetic field lines can penetrate the event horizon
 or they reach the outer-light cylinder.
 These cannot be necessary conditions for the Blandford--Znajek process
 to work.
 Moreover, the event horizon rather than the ergoregion 
 does not play an essential role in the
 energy extraction
 (this issue was addressed in, e.g., Refs.~\cite{Komissarov:2002dj,Ruiz:2012te} on
 the basis of numerical simulations).

 Generally speaking, the most important issue for extracting the rotational energy from
 black holes is angular-momentum transport to gain the energy in the ergoregion.
 The outward energy flux is just a by-product of 
 the angular-momentum transport. 
 In the Einstein gravity the spacetime metric can couple to energy--momentum
 tensors for any matters or fields. 
 If one wishes to extract the energy and angular momentum from the (black
 hole) spacetime, any other method does not exist except for the method
 via the energy--momentum tensor 
 (for gravitational waves, the energy--momentum pseudotensor).
 To elucidate an extraction mechanism of the rotational energy, 
 we should examine 
 what contents of the energy--momentum tensor mainly contribute to the
 angular-momentum transport 
 (the extraction mechanism for a general energy--momentum tensor is discussed in
 Ref.~\cite{Lasota:2013kia}).
 Such angular-momentum and energy transfer, i.e.,
 local conservation of the energy--momentum tensor, should be governed by
 local physics causally connected.
 Thus, the extraction process can be roughly dissected into three parts: 
 generating an energy by angular-momentum transfer in the
 ergoregion, 
 transporting the gained energy to a region far away from the black
 hole, and 
 disposing of garbage that has lost its angular momentum and energy to the
 black hole.
 The essence of the energy extraction mentioned previously is nothing but
 this generating process.
 On the other hand, global configurations of the magnetic fields at the
 event horizon and at a far region are related to the
 disposing process and the transporting one, respectively.

 After all, ``boundary conditions'' at the event horizon such as 
 whether the magnetic field lines can penetrate the horizon 
 cannot be a necessary condition for the energy extraction.
 In fact, even if the magnetic field lines failed to penetrate the horizon and
 could never get drawn into the black hole, the energy extraction by the
 Blandford--Znajek process can succeed.
 One may say that the energy is not extracted from any black hole in
 that case.
 However, the fact remains that the gravitational energy measured in an
 asymptotic region is extracted from the total system including the spacetime.
 Moreover, even though the magnetic field lines cannot reach the outer-light
 cylinder, we can say that the Blandford--Znajek process is at work if the
 magnetic field lines extend to a region sufficiently far away
 from the black hole and the
 gained energy in the ergoregion can be transferred there.

 In this paper, we show that the essential mechanism of the
 Blandford--Znajek process is determined by local physics
 in the neighborhood of the ergoregion.
 Of course, global configurations of the magnetic field and the electric current 
 are important to make the 
 Blandford--Znajek process efficiently and successfully sustaining.
 However, this is just a stage rather than a principal role.
 We stress that, in general, 
 how to extract rotational
 energy from the spacetime and 
 how to arrange appropriate configurations
 of the magnetic field or electric current for extracting the energy 
 are different questions.
 Furthermore, we should separately consider the kinematical properties
 without globally solving the equations of motion and dynamical properties. 

 The rest of the paper is organized as follows.
 We first review the field-sheet formalism for force-free electromagnetic
 fields.
 Then, we explicitly show a correspondence between the energy-extraction
 mechanisms by stationary, axisymmetric force-free electromagnetic
 fields and rigidly rotating strings. 
 In Appendix~\ref{app:string} 
 we briefly summarize some results for the rigidly rotating
 strings shown in Ref.~\cite{Kinoshita:2016lqd}.

 \section{Field-sheet formalism for force-free electromagnetic fields}

 In this section, in order to 
 elucidate 
 the similarity between Nambu--Goto
 strings and magnetic field lines of force-free fields, we will rewrite
 the equations of motion for force-free electromagnetic fields according
 to a ``field-sheet''
 formalism~\cite{Carter:1979,Uchida:1997a,Uchida:1997b,Gralla:2014yja}.
 Unless otherwise specified, Newton's constant $G_\mathrm{N}$ and the
 speed of light $c$ are set to unity hereafter.

Let $F_{\mu\nu}$ be 
the electromagnetic field strength 
and 
let $j_\mu$ be the current density four-vector of electric charge. 
In terms of $F_{\mu\nu}$, the Maxwell equations are given by
\begin{align}
\label{eq:Maxwell}
\nabla_\alpha F^{\mu \alpha}=4\pi j^\mu, \quad
\nabla_{[\:\!\mu}F_{\nu\lambda\:\!]}=0.
\end{align}
In general, when an electromagnetic field interacts with charged matter
such as plasma, 
the energy--momentum tensor for the electromagnetic field,
\begin{align}
T_{\mu\nu} \equiv 
 F_{\mu\alpha}F_\nu{}^\alpha-\frac{1}{4}\:\!F^{\alpha\beta}F_{\alpha\beta} \:\!g_{\mu\nu}, 
\end{align}
satisfies 
\begin{align}
\nabla_\mu T^\mu{}_\nu=-4\pi F_{\nu \alpha}\:\!j^\alpha, 
\end{align}
where the right-hand side of the above equation means the Lorentz force.
In the situation where the force density four-vector 
$F_{\nu\alpha}\:\!j^\alpha$ can be neglected, i.e., 
\begin{align}
\label{eq:ff}
F_{\mu\alpha}j^\alpha=0,
\end{align}
which is known as the 
{\em force-free condition}, 
the energy--momentum tensor of the electromagnetic field
is individually conserved $\nabla_\mu T^\mu{}_{\nu}=0$, 
so that neither angular momentum
nor energy is exchanged between the electromagnetic field and the other matter. 
The Maxwell equations~\eqref{eq:Maxwell} together with the force-free condition~\eqref{eq:ff} 
yield the equations
\begin{align}
F_{\mu\alpha}\nabla_\beta F^{\alpha \beta}=0, \quad
\nabla_{[\:\!\mu}F_{\nu\lambda\:\!]}=0.
\end{align}
The dynamics described by these equations is 
{\em force-free electrodynamics} (FFE). 

An important property for the force-free electromagnetic fields 
satisfies
\begin{equation}
 F_{\alpha\beta} {}^*\!F^{\alpha\beta}=0 ,
\end{equation}
where ${}^*\!F_{\mu\nu}$ is 
the dual of $F_{\mu\nu}$, defined by 
${}^*\!F_{\mu\nu} \equiv F^{\alpha \beta}\epsilon_{\alpha \beta\mu\nu}/2$.
These fields are called {\em degenerate}.
Note that the force-free condition for nonzero $j^\mu$ implies that
$F_{\mu\nu}$ is degenerate,%
\footnote{
In four dimensions, 
the relation 
$F_{[\alpha\beta}F_{\mu\nu]} = - \epsilon_{\alpha\beta\mu\nu} 
F_{\lambda\rho}{}^*\!F^{\lambda\rho}/12$ 
is satisfied.
While $F_{\mu\nu}j^\mu = 0$, we have 
$F_{\lambda\rho}{}^*\!F^{\lambda\rho}=0$ if $j^\mu$ is nonzero. 
} 
but degeneracy does not always lead to force-freeness.
To clarify the physical meanings, 
let $t^\mu$ be a four-velocity of a timelike observer.
Then, the electric and magnetic fields measured by this observer are given
by $\mathsf{E}^\mu=F^{\mu\nu}t_\nu$ and 
$\mathsf{B}^\mu=-{}^*\! F^{\mu\nu} t_\nu$,
respectively.
The degenerate condition physically means that 
$\mathsf{E}^\alpha \mathsf{B}_\alpha =0$,
where this relation holds even for an arbitrary observer $t^\mu$
because $F_{\alpha\beta} {}^*\!F^{\alpha\beta}$ is scalar. 
Now, the fact that $F_{\mu\nu}$ is closed together with the degeneracy implies that 
${}^*\!F_{\mu\nu}$ is tangent to a 2D submanifold, $\mathscr{S}$.
Furthermore, we assume that 
$F_{\mu\nu}$ is {\em magnetically dominated}, i.e., 
\begin{align}
F_{\alpha\beta}F^{\alpha\beta}
=2\:\!(\mathsf{B}^\alpha \mathsf{B}_\alpha
-\mathsf{E}^\alpha \mathsf{E}_\alpha)>0.  
\end{align}
Naively, this condition implies that the magnetic field should be stronger
than the electric field.
Because this condition is also described by a scalar, there is a notion
independent of observers.
The degeneracy and magnetically dominated condition for $F_{\mu\nu}$ guarantee 
the existence of a pure magnetic frame in which $\mathsf{E}^\mu=0$.
Therefore, $F_{\alpha\beta}F^{\alpha\beta} > 0$ states that 
the electromagnetic field $F_{\mu\nu}$ 
is purely magnetic with its magnitude defined
by 
\begin{align}
B \equiv \sqrt{\frac{F_{\mu\nu}F^{\mu\nu}}{2}}.
\end{align}
Namely, we can take ${}^*\!F_{\mu\nu}$ in the form 
\begin{equation}
 {}^*\!F_{\mu\nu} = B \sigma_{\mu\nu}, 
  \label{eq:magnetic_twoform}
\end{equation} 
where $\sigma_{\mu\nu}$ denotes the 2D volume element on
$\mathscr{S}$.
It turns out that the magnetically dominated condition for $F_{\mu\nu}$
is equivalent to $\sigma_{\mu\nu}\sigma^{\mu\nu} = -2$, and then 
$\mathscr{S}$ becomes timelike. 
We call such $\mathscr{S}$ a
{\em field sheet}. 
The scalar function $B(>0)$ irrelevant to coordinate systems 
means the proper magnitude of the
magnetic field, which can be observed at the rest frame of the magnetic
field lines.
Moreover, the magnetic tension and pressure are given by
this quantity, so that they are also proper quantities independent of
coordinate systems.

We describe the dynamics of field sheets as string world sheets. 
In terms of ${}^*\!F_{\mu\nu}$, the equations of FFE are rewritten as 
\begin{align}
{}^*\!F^{\alpha\beta}\:\!\nabla_{[\:\!\mu} {}^*\!F_{\alpha\beta\:\!]}=0,
\quad
\nabla_\alpha {}^*\!F^{\alpha\mu}=0.
\end{align}
Substituting Eq.~(\ref{eq:magnetic_twoform}) into the above, 
we obtain the equations of FFE
in terms of $B$ and $\sigma_{\mu\nu}$ given by 
\begin{align}
\label{eq:EqSigma1}
&\sigma^{\alpha\beta}\nabla_\alpha \sigma_{\beta\mu}
=N_\mu{}^{\alpha}\:\!\nabla_\alpha \ln B,
\\
\label{eq:EqSigma2}
&\nabla_\alpha (B \:\!\sigma^{\alpha \mu})=0, 
\end{align}
where $N_{\mu}{}^{\nu}$ is defined by 
\begin{align}
&N_\mu{}^{\nu} \equiv g_\mu{}^{\nu}-h_\mu{}^{\nu}, 
\\
&\ h_\mu{}^{\nu} \equiv \sigma_{\mu\alpha}\:\!\sigma^{\alpha\nu}. 
\end{align}
Note that $h_{\mu\nu}$ is the induced metric on the field sheet and
$N_{\mu}{}^{\nu}$ is the projection tensor onto directions normal to the
field sheet.
The former equation (\ref{eq:EqSigma1}) describes dynamics of field
sheets acted on by a force associated with $B$, i.e., magnetic pressure;
the latter equation (\ref{eq:EqSigma2}) does conservation of the
magnetic flux.
This expression tells us that the dynamics of field sheets in FFE is similar
to the dynamics of world sheets for Nambu--Goto strings.
In fact, it is known that 
the equations of motion for a Nambu--Goto string can be written
as 
\begin{equation}
 \sigma^{\alpha\beta}\nabla_\alpha \sigma_{\beta\mu} = 0 ,
\end{equation}
where $\sigma_{\mu\nu}$ denotes the volume element of the world sheet in
this case~\cite{Schild:1976vq}.
Comparing the above equation with Eqs.~(\ref{eq:EqSigma1}) and
(\ref{eq:EqSigma2}), we can easily notice that the only difference
between the dynamics of field sheets and world sheets is the addition of
the extra scalar
quantity $B$, which describes the magnetic tension and pressure.
(In the flat spacetime such a correspondence was discussed in
Refs.~\cite{Kastrup:1979fz,Rinke:1979mw,Nambu:1981gt}.)

From a geometrical point of view, their meanings are so clear.
In general, the extrinsic curvature of a submanifold is defined by 
\begin{equation}
 K^\lambda{}_{\mu\nu} \equiv - h_\mu{}^\alpha h_\nu{}^\beta 
  \nabla_\beta h_\alpha{}^\lambda ,
\end{equation}
where $h_{\mu\nu}$ is the induced metric on the submanifold.
If the submanifold is 2D, we have 
\begin{equation}
 K^\lambda{}_{\mu\nu} h^{\mu\nu} = 
  - \sigma^{\alpha\beta}\nabla_\alpha \sigma_{\beta}{}^\lambda .
\end{equation}
Thus, the dynamics of field sheets and world sheets belong to the same
class of dynamics of a 2D surface 
(see Ref.~\cite{Carter:2000wv} and references therein). 
Because every degenerate, closed two-form defines a foliation of
spacetime,%
\footnote{See Appendix A in Ref.~\cite{Compere:2016xwa}.}
the field sheets represented by $F_{\mu\nu}$ can define a
2D timelike foliation with a coordinate transformation
$x^\mu = X^\mu(\tau,\sigma,\alpha,\beta)$, where $(\tau,\sigma)$ denote
local coordinates on field sheets $\mathscr{S}$ 
defined by $\alpha=\mathrm{const.}$
and $\beta=\mathrm{const}$.
Note that, once $\alpha$ and $\beta$ are fixed, $X^\mu$ give embedding
functions of $\mathscr{S}$.
Let $\partial_a$ be coordinate derivatives with respect to 2D local
coordinates on the field sheet.
The Latin indices denote intrinsic components on the field sheet.
The volume element $\sigma_{\mu\nu}$ is related to $X^\mu$ as 
\begin{align}
\sigma^{\mu\nu}
=\sigma^{ab}h_a{}^\mu h_b{}^\nu,
\end{align} 
where $h_a{}^\mu \equiv \partial_a X^\mu$, and 
$\sigma_{ab}$ 
is the intrinsic
volume element on $\mathscr{S}$. 
Moreover, the induced metric can be intrinsically written as 
\begin{equation}
 h_{ab} = g_{\mu\nu}(X) h_a{}^\mu h_b{}^\nu
  = h_{\mu\nu} h_a{}^\mu h_b{}^\nu . 
\end{equation}
By using these intrinsic quantities, we can
rewrite Eq.~(\ref{eq:EqSigma1}) in terms of $X^\mu$ as 
\begin{equation}
 D^2 X^\mu + \Gamma^\mu{}_{\alpha\beta}D_a X^\alpha D^a X^\beta
  = N^{\mu\alpha}\partial_\alpha \ln B ,
\end{equation}
where $D_a$ denotes the covariant derivative with respect to $h_{ab}$
and $\Gamma^\mu{}_{\alpha\beta}$ is the Christoffel symbol associated
with $g_{\mu\nu}$.
Note that, intriguingly, this equation is derived from the following action: 
$S[X^\mu] = - \int d^2\sigma B(X)\sqrt{-h}$,
where $h$ denotes the determinant of the induced metric $h_{ab}$.

 It is instructive 
 to compare a perfect fluid in terms of
 conservation of the energy--momentum tensor.
 As is well known, the energy--momentum tensor of a perfect fluid
 is given by 
 \begin{equation}
  T^{\mu\nu} = \rho u^\mu u^\nu + p (g^{\mu\nu} + u^\mu u^\nu) ,
 \end{equation}
 where $u^\mu$ is a four-velocity of the fluid, and 
 $\rho$ and $p$ are 
 the energy density and pressure 
 in its rest frame, respectively.
 Decomposing the conservation law $\nabla_\mu T^{\mu\nu} = 0$ into
 tangential components and normal ones with respect to $u^\mu$ yields 
 \begin{equation}
  \begin{aligned}
   (\rho + p)u^\mu\nabla_\mu u^\nu &= 
   - (g^{\mu\nu} + u^\mu u^\nu)\nabla_\mu p ,\\
   u^\mu \nabla_\mu \rho + (\rho + p)\nabla_\mu u^\mu &= 0.
  \end{aligned}
 \end{equation}
 If the fluid is pressureless (namely, a dust fluid), the four-velocity
 of the fluid obeys geodesic equations.
 In a parallel manner, we consider an energy--momentum tensor given by 
 \begin{equation}
  \begin{aligned}
   T^{\mu\nu} =& - \mu h^{\mu\nu} + \tilde{p}(g^{\mu\nu}-h^{\mu\nu}) \\
   =& (\mu + \tilde{p}) \sigma^{\mu\alpha}\sigma^{\nu}{}_\alpha
   + \tilde{p} g^{\mu\nu} ,  
  \end{aligned}
 \end{equation}
 where $\mu$ is a tension equal to its energy density and $\tilde{p}$ is
 a normal pressure.
 Note that Nambu--Goto strings have a constant $\mu$ and $\tilde{p}=0$, and
 magnetically dominated electromagnetic fields have 
 $\mu = \tilde{p} = B^2/2$.
 Decomposing the energy--momentum conservation into components tangential
 and normal to the field sheet yields 
 \begin{equation}
  \begin{aligned}
   (\mu + \tilde{p})\sigma^{\alpha\beta}\nabla_\alpha \sigma_{\beta\mu}
   &= (g_\mu{}^\alpha - h_\mu{}^\alpha)\nabla_\alpha \tilde{p} ,\\
  \sigma^{\alpha\mu} \nabla_\alpha \mu+(\mu+\tilde{p}) h^\mu{}_\nu
   \nabla_\alpha \sigma^{\alpha\nu}&=0 .
  \end{aligned}
 \end{equation}

 Thus, it turns out that the world line and its volume element $u_\mu$ for a
 fluid element of perfect fluids correspond to the world sheet (field
 sheet) and its volume element $\sigma_{\mu\nu}$ for a Nambu--Goto string
 or a magnetic field line.
 Moreover, in each pressureless case the world lines of dusts become
 geodesics and the world sheets of Nambu--Goto strings are extremal
 surfaces.

 If the system admits some symmetries and such a symmetry is
 characterized by a Killing vector field $\xi^\mu$, the conservation law
 of the energy--momentum tensor in terms of $\xi^\mu$ yields
 \begin{equation}
  \begin{aligned}
   0 &= \nabla_\mu(T^{\mu\nu}\xi_\nu) \\
   &= - \nabla_\mu [(\mu + \tilde{p})h^{\mu\nu}\xi_\nu]
   + \xi^\mu \nabla_\mu \tilde{p} \\
   &= - \nabla_\mu [(\mu + \tilde{p})h^{\mu\nu}\xi_\nu] ,
  \end{aligned}
 \end{equation}
 where we have used $\xi^\mu\nabla_\mu \tilde{p} = 0$ because of the
 symmetry.
 This conservation law is equivalent to that of a string with an
 effective tension $\mu + \tilde{p}$.
 Thus, even though the pressure does not vanish, $\tilde{p}\neq 0$,
 kinematic properties such as the conservation law can be identical thanks
 to the symmetry.

 \section{Energy extraction by force-free electromagnetic fields}
 \label{sec:3}
 
  \subsection{Stationary and axisymmetric force-free electromagnetic fields}
  
  In this section, we consider stationary and axisymmetric force-free
  electromagnetic fields in a rotating black hole, and reinterpret 
  energy-extraction processes via the force-free electromagnetic fields
  from a perspective of the field-sheet formalism.
  Such stationary and axisymmetric force-free electromagnetic fields in
  a stationary and axisymmetric spacetime have been comprehensively 
  investigated in the literature, and their basic properties are 
  well known~\cite{Frolov:1998wf}.
  Note that the above assumptions are conventional ones when the
  Blandford--Znajek process has been discussed. 
  In the field-sheet approach, we will follow 
  Ref.~\cite{Gralla:2014yja}.

  In order that we will later focus on the Kerr spacetime as an explicit
  example, we suppose that the spacetime is stationary and axisymmetric
  and admits two Killing vector fields $(\partial_t)^\mu$ and
  $(\partial_\phi)^\mu$, which respectively represent time translation
  symmetry and axisymmetry of the spacetime as 
  $\mathcal{L}_{\partial_t}g_{\mu\nu} =
  \mathcal{L}_{\partial_\phi}g_{\mu\nu} = 0$.
  Here, $\mathcal{L}$ denotes the Lie derivative.
  Its metric can be written as 
  \begin{equation}
   g_{\mu\nu} dx^\mu dx^\nu
    = g_{tt}dt^2 + 2g_{t\phi}dtd\phi + g_{\phi\phi}d\phi^2
    + g_{rr}dr^2 + g_{\theta\theta}d\theta^2 ,
    \label{eq:metric}
  \end{equation}
  where we assume that $g_{tt}g_{\phi\phi} - g_{t\phi}^2 = 0$ 
  characterizes the event horizon and 
  $g_{tt}g_{\phi\phi} - g_{t\phi}^2 < 0$ implies 
  the outside of the black hole.
  Here, we have taken a coordinate system in which the Killing vector
  fields $(\partial_t)^\mu$ and $(\partial_\phi)^\mu$ are manifestly
  orthogonal to 2D surfaces spanned by $(r,\theta)$ such as
  the Boyer--Lindquist coordinates for the Kerr spacetime. 
  Moreover, we suppose that 
  the electromagnetic fields share the same Killing symmetries such
  that 
  $\mathcal{L}_{\partial_t}F_{\mu\nu} = \mathcal{L}_{\partial_\phi}F_{\mu\nu} = 0$.

  It is known that 
  stationary and axisymmetric force-free electromagnetic fields are
  given by the following field strength: 
  \begin{equation}
   F = \frac{1}{2}F_{\mu\nu}dx^\mu \wedge dx^\nu
    = d\psi \wedge (\eta - d\varphi) ,
    \label{eq:SAFFE_fieldstrength}
  \end{equation}
  where $\eta \equiv d\phi - \omega(\psi)dt$, 
  and
  both $\psi$ and $\varphi$ are scalar functions independent of $t$
  and $\phi$.
  This yields a common component representation of the field strength:
  \begin{equation} 
   F_{\mu t} = F_{\mu\nu}(\partial_t)^\nu 
    = - \omega(\psi)\partial_\mu \psi ,\quad    
   F_{\mu \phi} = F_{\mu\nu}(\partial_\phi)^\nu 
    = \partial_\mu \psi ,\quad
    F_{r\theta} = \frac{\partial\varphi}{\partial r}
    \frac{\partial\psi}{\partial\theta}
    - \frac{\partial\varphi}{\partial\theta}
    \frac{\partial\psi}{\partial r} ,
  \end{equation}
  which are related to an electric field, and poloidal and toroidal
  components of the magnetic field, respectively.
  Note that we can easily confirm that $dF=0$ is satisfied.
  The form of this field strength (\ref{eq:SAFFE_fieldstrength}) can be
  expressed as $F=d\psi\wedge d\hat{\phi}$ in terms of the 
  so-called Euler potentials given by the following scalar functions: 
  $\psi$ and $\hat{\phi} \equiv \phi - \omega(\psi) t - \varphi$
  (see, e.g., Ref.~\cite{Gralla:2014yja}).
  This means that the field sheets are represented by the intersections of the
  hypersurfaces of constant $\psi$ and $\hat{\phi}$.
  Each magnetic field line given by the field sheets 
  is rotating with angular velocity $\omega(\psi)$. 
  Obviously, the field sheets described by the same $\psi$ have the same
  angular velocity because of the axisymmetry.
  Such a $\psi$-constant surface is a so-called magnetic surface, and
  $2\pi\psi = \int F$ gives the magnetic flux across the region
  enclosed by the magnetic surface.

  The ``proper'' magnetic field $B$ is given by  
  \begin{equation}
  B^2 \equiv \frac{1}{2}F_{\mu\nu}F^{\mu\nu}
   = \eta^2 |\nabla\psi|^2 + |\nabla\psi|^2|\nabla\varphi|^2
   - (\nabla\psi \cdot \nabla\varphi)^2.
   \label{eq:B_squared}
  \end{equation}
  Here, we have written the contractions 
  for vector fields $u$ and $v$ as 
  $u\cdot v = u_\mu v^\mu$ and $u^2 = u_\mu u^\mu$ in the abbreviated notation.
  In addition, we have 
  \begin{equation}
   F_{r\theta}F^{r\theta} = g^{rr}g^{\theta\theta}F_{r\theta}^2
    = |\nabla\psi|^2|\nabla\varphi|^2
    - (\nabla\psi \cdot \nabla\varphi)^2.
     \label{eq:Frtheta_squared}
  \end{equation}
  As we mentioned in the previous section, $B>0$ is a scalar function
  irrelevant to coordinate systems, which gives the magnetic tension and
  pressure.

  Now, $\chi$ is a corotating vector defined by 
  \begin{equation}
   \chi^\mu = \left(\frac{\partial}{\partial t}\right)^\mu
    + \omega(\psi)\left(\frac{\partial}{\partial \phi}\right)^\mu,
  \end{equation}
  which satisfies $\chi^\mu\eta_\mu = 0$ and 
  $\chi^2 = g_{tt} + 2g_{t\phi}\omega + g_{\phi\phi}\omega^2 
  = - (g_{t\phi}^2 - g_{tt}g_{\phi\phi})\eta^2$.
  Since the corotating vector $\chi$ satisfies $F_{\mu\nu}\chi^\nu = 0$,
  $\chi$ is tangential to the field sheet.
  It turns out that $\chi$ represents a corotating frame with the angular
  velocity $\omega(\psi)$ for each field sheet labeled by constant $\psi$.
  If $\chi$ (also, $\eta$) becomes null, the velocity of such a corotating
  frame has reached the speed of light.
  The locus characterized by $\chi^2 = 0$ is called the light surface.%
  \footnote{
  In general, two light surfaces can exist in a black hole
  spacetime: one is the inner-light surface near the black hole, and the
  other is the outer-light surface (light cylinder) in the far region.
  Since the inner-light surface is important for extracting the
  rotational energy from the black hole~\cite{Kinoshita:2016lqd}, 
  we will focus only on the inner-light surface hereafter.}
  Meanwhile, because a field sheet is a 2D surface, 
  there is another tangential vector linearly independent of $\chi$.
  We will introduce the other tangential vector as 
  \begin{equation}
   \begin{aligned}
    \lambda^\mu &\equiv {}^*\!F^{\mu\nu}\nabla_\nu t \\
    &= \frac{1}{\sqrt{-g}}
    \left[F_{\theta\phi}\left(\frac{\partial}{\partial r}\right)^\mu
    + F_{\phi r}\left(\frac{\partial}{\partial \theta}\right)^\mu
    + F_{r\theta}\left(\frac{\partial}{\partial \phi}\right)^\mu
    \right] .
   \end{aligned}
  \end{equation}
  By construction, this vector is tangential to the field sheet and
  normal to $\nabla_\mu t$, so that $\lambda$ becomes a generator of the
  intersection of the field sheet and a constant-$t$ surface.
  This means that the integral curves of $\lambda$ are magnetic field lines
  at a time $t$.
  The direction of
  $\lambda$ is radially outward for 
  $F_{\theta\phi}=\partial_\theta\psi > 0$, while its direction is
  radially inward for $\partial_\theta\psi < 0$.
  It turns out that the direction of $\lambda$ corresponds to that of
  the magnetic field.
  Note that, since $[\lambda,\chi]=\mathcal{L}_\lambda \chi = 0$ are
  satisfied, $\chi$ and $\lambda$ can be coordinate bases on the
  field sheet as $\chi=\partial/\partial\tau$ and 
  $\lambda=\partial/\partial\sigma$, where $\tau$ and $\sigma$ denote
  local coordinates on the field sheet.
  The dual 
  of $F_{\mu\nu}$ 
  is expressed as 
  \begin{equation}
   {}^*\!F^{\mu\nu} = \lambda^\mu \chi^\nu - \chi^\mu \lambda^\nu .
  \end{equation}
  This expression manifestly shows that it is a two-form tangential to the
  field sheet, i.e., proportional to the volume form of the
  field sheet as ${}^*\!F_{\mu\nu} = B\sigma_{\mu\nu}$.

  \subsection{Induced geometry on the field sheet}

  As we mentioned, $\chi$ and $\lambda$ can constitute the tangential 
  coordinate bases on the
  field sheet, so that components of the induced metric on the
  field sheet in terms of the coordinate system $(\tau,\sigma)$ 
  are given by 
  \begin{equation}
   \begin{aligned}
    h_{\tau\tau} &\equiv \chi^\mu\chi^\nu g_{\mu\nu}
    = g_{tt} + 2g_{t\phi}\omega + g_{\phi\phi}\omega^2 , \\
    h_{\tau\sigma} &\equiv \chi^\mu\lambda^\nu g_{\mu\nu}
    = \frac{F_{r\theta}}{\sqrt{-g}}(g_{t\phi} + g_{\phi\phi}\omega) ,\\
    h_{\sigma\sigma} &\equiv \lambda^\mu\lambda^\nu g_{\mu\nu}
    = \frac{1}{g_{t\phi}^2 - g_{tt}g_{\phi\phi}}|\nabla\psi|^2
    + g_{\phi\phi} \left(\frac{F_{r\theta}}{\sqrt{-g}}\right)^2 .
   \end{aligned}
  \end{equation}
  Since $\chi$ is the corotating vector, the light surface is
  characterized by $\chi^2 =h_{\tau\tau}= 0$.
  In addition, $\chi$ is the Killing vector with respect to the induced
  metric on the field sheet,%
  \footnote{ Each component of the induced metric is a function of only 
  $r$ and $\theta$.
  Since $\mathcal{L}_\chi r = \mathcal{L}_\chi \theta = 0$, we have 
  $\mathcal{L}_\chi h_{ab} = 0$ on the field sheet. 
  }
  so that the light surface corresponds to
  the Killing horizon on the field sheet. 
  This Killing horizon works as a causal boundary for various phenomena governed
  by the induced metric~\cite{Gralla:2014yja}.
  We find the following correspondence between the induced metric of 
  the field sheet and that of the rigidly
  rotating string (\ref{eq:string_hab}): 
  \begin{equation}
   \frac{F_{r\theta}}{\sqrt{-g}} \leftrightarrow \varphi', \quad
    \frac{1}{g_{t\phi}^2 - g_{tt}g_{\phi\phi}}|\nabla\psi|^2
    \leftrightarrow
    g_{rr}r'^2 + g_{\theta\theta}\theta'^2 .
  \end{equation}
  Indeed, we can explicitly show 
  \begin{equation}
   \mathcal{L}_\lambda \varphi 
    = \lambda^\mu \partial_\mu \varphi
    = \frac{1}{\sqrt{-g}}
    \left(\frac{\partial\psi}{\partial \theta}
     \frac{\partial\varphi}{\partial r}
     - \frac{\partial\psi}{\partial r}
     \frac{\partial\varphi}{\partial \theta}
    \right)
    = \frac{F_{r\theta}}{\sqrt{-g}} ,
    \label{eq:dvarphi}
  \end{equation}
  and 
  \begin{equation}
   \mathcal{L}_\lambda r =
    \lambda^\mu\partial_\mu r = 
    \frac{\partial_\theta \psi}{\sqrt{-g}},\quad
    \mathcal{L}_\lambda \theta =
    \lambda^\mu\partial_\mu \theta = -
    \frac{\partial_r \psi}{\sqrt{-g}} .
    \label{eq:dr_dtheta}
  \end{equation}
  If we introduce the local coordinates as $\chi=\partial/\partial\tau$ 
  and $\lambda=\partial/\partial\sigma$ restricted on the field sheet, 
  the correspondence becomes more apparent.
  Thus, the intrinsic geometry on the field sheet of the stationary,
  axisymmetric force-free electromagnetic field and that on the world
  sheet of the rigidly rotating string have entirely identical properties.

  \subsection{Angular-momentum flux and energy flux}
  
  Since the spacetime has time-translational symmetry and axisymmetry,
  we have conservation laws for the energy--momentum tensor that 
  are associated with each symmetry.
  In particular, the energy--momentum tensor of the force-free electromagnetic
  field is conserved independently of other matter.
  The angular-momentum conservation 
  $\nabla_\mu(T^\mu{}_\nu(\partial_\phi)^\nu) = 0$ 
  yields 
  \begin{equation}
   \begin{aligned}
    0 &= \frac{1}{\sqrt{-g}}
    \partial_\mu (\sqrt{-g}T^\mu{}_\nu(\partial_\phi)^\nu) \\
    &= - \frac{1}{\sqrt{-g}}
    \partial_\mu (\sqrt{-g}F^{\mu\nu}\partial_\nu \psi)\\
    &= - \frac{1}{\sqrt{-g}}
    \left(\partial_\theta \psi \frac{\partial}{\partial r}
    - \partial_r \psi \frac{\partial}{\partial \theta}\right)
    \sqrt{-g}F^{r\theta} .
   \end{aligned}
  \end{equation}
  This indicates that $\sqrt{-g}F^{r\theta}$ should depend only on
  $\psi$ as 
  \begin{equation}
   I(\psi) \equiv \sqrt{-g} F^{r\theta} .
  \end{equation}
  Note that this scalar function%
  \footnote{ 
  Since 
  $I(\psi) = {}^*F_{\mu\nu}(\partial_t)^\mu(\partial_\phi)^\nu$, 
  we can see that $I(\psi)$ is scalar. 
  }
  is constant on the field sheets, 
  and it is one of
  the characteristic quantities characterizing  
  the stationary and axisymmetric 
  force-free electromagnetic field as well as $\psi$ and $\omega(\psi)$.
  By using $\lambda$, this angular-momentum conservation can be
  rewritten as 
  \begin{equation}
   \frac{1}{\sqrt{-g}}
    \partial_t (\sqrt{-g}T^t{}_\nu(\partial_\phi)^\nu)
    = \lambda^\mu \partial_\mu I(\psi) .
    \label{eq:local_angular_momentum_conservation}
  \end{equation}
  Therefore, 
  $I(\psi)$ is identified with the angular-momentum flux 
  (per unit magnetic flux $d\psi$)
  flowing on the field sheet, which should be conserved for each field sheet.
  Note that $-I(\psi)$ gives the outward angular-momentum flux for 
  $\partial_\theta \psi >0$, while $I(\psi)$ is the outward flux for 
  $\partial_\theta \psi <0$.
  This is because the direction of $\lambda$ changes depending on
  the sign of $\partial_\theta \psi$ (for instance, recall that 
  $\mathcal{L}_\lambda r = \partial_\theta \psi /\sqrt{-g}$ in 
  Eq.~(\ref{eq:dr_dtheta})).
  It is well known that $I(\psi)$ is connected with the electric current
  according to Amp\`ere's law. 
  However, we stress that this is {\em electromagnetic} angular-momentum flux
  without involving matter such as charged particles, 
  because the relevant energy--momentum tensor consists only of the
  electromagnetic field and it is individually conserved thanks to the
  force-free condition. 
  Similarly, the energy conservation 
  $\nabla_\mu(T^\mu{}_\nu(\partial_t)^\nu) = 0$ leads to 
  \begin{equation}
   \frac{1}{\sqrt{-g}}
    \partial_t (-\sqrt{-g}T^t{}_\nu(\partial_t)^\nu)
    = \lambda^\mu \partial_\mu[\omega(\psi)I(\psi)] ,
    \label{eq:local_energy_conservation}
  \end{equation}
  where $\omega(\psi)I(\psi)$ is identified with the conserved energy flux
  flowing on the field sheet.
  As in the case of the angular-momentum flux, 
  $-\omega(\psi)I(\psi)$ gives the outward energy flux for 
  $\partial_\theta \psi >0$, 
  while $\omega(\psi)I(\psi)$ is the outward flux for 
  $\partial_\theta \psi <0$.
  Integrating Eqs.~(\ref{eq:local_angular_momentum_conservation}) and
  (\ref{eq:local_energy_conservation}) over a 
  3D volume on a $t$-constant hypersurface  
  surrounding the
  black hole, we have total fluxes of angular
  momentum $\mathcal{J}$ and energy $\mathcal{M}$, 
  i.e., each extraction rate from the black hole,%
  \footnote{We denote the mass and angular momentum of the black hole
  as $\mathcal{M}_\mathrm{BH}$ and $\mathcal{J}_\mathrm{BH}$.
  Each conservation law yields   
  $\frac{d\mathcal{M}}{dt} + \frac{d\mathcal{M}_\mathrm{BH}}{dt} = 0$
  and 
  $\frac{d\mathcal{J}}{dt} + \frac{d\mathcal{J}_\mathrm{BH}}{dt} = 0$.
  }
  as 
  \begin{equation}
   \frac{d\mathcal{J}}{dt} = - 2\pi \int I(\psi)d\psi ,\quad
    \frac{d\mathcal{M}}{dt} = - 2\pi \int \omega(\psi)I(\psi)d\psi .
    \label{eq:total_fluxes}
  \end{equation}

  Now, we find from Eq.~(\ref{eq:dvarphi}) that 
  \begin{equation}
   I(\psi) = \sqrt{-g}F^{r\theta}
    = (-g)g^{rr}g^{\theta\theta}\frac{F_{r\theta}}{\sqrt{-g}}
    = (g_{t\phi}^2 - g_{tt}g_{\phi\phi})\mathcal{L}_\lambda \varphi ,
  \end{equation}
  and also we find from Eqs.~(\ref{eq:B_squared}) and \eqref{eq:Frtheta_squared} that 
  \begin{equation}
   \begin{aligned}
    I(\psi)^2 &= (g_{t\phi}^2 - g_{tt}g_{\phi\phi})
    (B^2 - \eta^2 |\nabla\psi|^2) \\
    &= (g_{t\phi}^2 - g_{tt}g_{\phi\phi})B^2 + \chi^2 |\nabla\psi|^2 .
   \end{aligned}
   \label{eq:I2_B2_relation}
  \end{equation}
  Combining the above two results, we can obtain 
  \begin{equation}
   \hat{q} \equiv \mp \frac{I(\psi)}{B} 
    = \mp \frac{(g_{t\phi}^2 - g_{tt}g_{\phi\phi})\mathcal{L}_\lambda \varphi}
    {\sqrt{(g_{t\phi}^2 - g_{tt}g_{\phi\phi})(\mathcal{L}_\lambda
    \varphi)^2
    - \chi^2 [g_{rr}(\mathcal{L}_\lambda r)^2 + 
    g_{\theta\theta}(\mathcal{L}_\lambda \theta)^2]}} ,
    \label{eq:electromagnetic_q}
  \end{equation}
  where we take the upper minus sign for $\partial_\theta \psi > 0$ and 
  the lower plus sign for $\partial_\theta \psi < 0$.
  In a unified manner, alternatively, it can be rewritten as 
  \begin{equation}
   \hat{q} = 
    - \frac{(g_{t\phi}^2 - g_{tt}g_{\phi\phi})d\varphi/dr}
    {\sqrt{(g_{t\phi}^2 - g_{tt}g_{\phi\phi})(d\varphi/dr)^2
    - \chi^2 [g_{rr} + g_{\theta\theta}(d\theta/dr)^2]}} ,
    \label{eq:electromagnetic_q_alt}
  \end{equation}
  where 
  $d\varphi/dr \equiv \mathcal{L}_\lambda \varphi/\mathcal{L}_\lambda r$
  and 
  $d\theta/dr \equiv \mathcal{L}_\lambda \theta/\mathcal{L}_\lambda r$.
  This quantity $\hat{q}$ means the specific angular-momentum flux per
  unit tension, and its expression is identical to that of the specific
  angular-momentum flux for the rigidly rotating
  strings~\cite{Kinoshita:2016lqd}
  (see also Eq.~(\ref{eq:string_q})).
  Similarly, $\omega(\psi)\hat{q}$ is the specific energy flux per unit
  tension.
  Since the sign of $\hat{q}$ has been defined such that $\hat{q}$ should be
  positive if a radially outward flux, $\hat{q} > 0$ and 
  $\omega\hat{q} > 0$ mean angular
  momentum extracting and energy extracting, respectively.
  It is worth noting that the sign of $\hat{q}$ is irrelevant to the
  direction of $\lambda$ as is clear from Eq.~(\ref{eq:electromagnetic_q_alt}).
  In other words, whether it is an extracting process or an injecting one does not
  depend on the direction of the magnetic field but the configuration of
  the magnetic field line.
  One thing to keep in mind as a difference from the
  cases of the rigidly rotating strings is that the specific 
  angular-momentum flux $\hat{q}$ is not
  conserved while $I(\psi)$ is conserved on the field sheet.
  The tension of Nambu--Goto strings is constant, while the magnetic
  tension associated with $B$ can vary even on the field sheet.
  Therefore, the specific angular-momentum flux per tension should be
  not necessarily conserved, or more accurately the scalar function $B$
  can work as an effective tension for the field sheet.%
  \footnote{
  As can be seen from the energy--momentum tensor, $B^2$ describes the magnetic
  tension per unit area, which has the same dimension as pressure.
  Now, $B$ describes the magnetic tension per unit magnetic flux.
  }
  
  So far we have shown that the intrinsic structures on the field sheet
  are identical to those on the world sheet of the rigidly rotating
  string.
  This does not mean that the global, extrinsic structures of both objects
  will be identical.
  The equations of motion for each are indeed similar, but they are not 
  identical, as seen in the previous section.
  In general, the global configurations of the Nambu--Goto 
  strings and the magnetic field
  lines are different, and 
  such global configurations should be determined by each dynamics by 
  solving the equations of motion.
  However, it is noteworthy that kinetic properties can be
  locally determined without solving the equations of motion. 
  In what follows we will see that the specific angular-momentum flux is
  in fact governed by local relations on the light surface.
  On the light surface, which is characterized by 
  \begin{equation}
   \chi^2 = g_{tt} + 2 g_{t\phi}\omega(\psi) + g_{\phi\phi}\omega(\psi)^2
    = 0,
    \label{eq:electromagnetic_chi2}
  \end{equation}
  we have the specific angular-momentum flux 
  \begin{equation}
   \hat{q}^2 = (g_{t\phi}^2 - g_{tt}g_{\phi\phi})|_{\chi^2 = 0} ,
    \label{eq:electromagnetic_q2}
  \end{equation}
  by substituting $\chi^2 = 0$ into Eq.~(\ref{eq:I2_B2_relation}).
  It turns out that 
  this expression is identical to that of the specific angular-momentum
  flux for the rigidly rotating strings (\ref{eq:string_q2}).
  Since Eqs.~(\ref{eq:electromagnetic_chi2}) and
  (\ref{eq:electromagnetic_q2}) are equations in terms of $r$ and
  $\theta$ essentially, 
  they give relations among the angular velocity $\omega$, 
  the specific angular-momentum flux $\hat{q}$, and the locus of the
  light surface 
  $(r_\mathrm{LS}, \theta_\mathrm{LS})$.
  Thus, we conclude that both the stationary, axisymmetric force-free
  electromagnetic fields and the rigidly rotating strings have 
  identical relations.
  It is worth noting that these relations are determined by the
  background spacetime metric irrelevant to the dynamics of the
  electromagnetic field. 
  Even though the dynamics of electromagnetic fields and rigidly
  rotating strings are different, i.e., their global configurations
  are different, the same locus of the light surface provides the same
  angular velocity and specific angular-momentum flux. 
  This fact implies that the specific angular-momentum flux is
  kinematically determined and both 
  extracting mechanisms of energy and angular momentum via the force-free
  electromagnetic fields and the rigidly rotating strings are
  essentially identical.

  \subsection{Energy extraction in the Kerr spacetime}
  
  Now, to discuss the energy-extraction process from a rotating black
  hole explicitly, let us focus on the Kerr spacetime.
  Since the major properties of the force-free
  electromagnetic fields and the rigidly rotating strings are identical,
  as seen, most
  of the following argument and its results are the same as those shown in 
  Ref.~\cite{Kinoshita:2016lqd}. 
  For more details, refer to it.

  In the Boyer--Lindquist coordinates the metric of the Kerr spacetime
  with mass $M$ and angular momentum $aM$ is given by 
  \begin{equation}
   g_{\mu\nu}dx^\mu dx^\nu = 
    - dt^2 + \frac{2Mr}{\Sigma}(dt - a\sin^2\theta d\phi)^2
    + \frac{\Sigma}{\Delta}dr^2 + \Sigma d\theta^2 
    + (r^2 + a^2)\sin^2\theta d\phi^2 ,
  \end{equation}
  where 
  \begin{equation}
   \Sigma(r,\theta) = r^2 + a^2\cos^2\theta, \quad \Delta(r) = r^2 + a^2 - 2Mr .
  \end{equation}
  For simplicity and without loss of generality, we assume $a>0$.
  This metric gives $g_{t\phi}^2 - g_{tt}g_{\phi\phi} = \Delta \sin^2\theta$, and 
  the event horizon lies at $r=r_\mathrm{h}\equiv M+\sqrt{M^2 - a^2}$
  defined by $\Delta(r_\mathrm{h}) = 0$.
  Note that we have $\Delta > 0$ outside the black hole $r>r_\mathrm{h}$.
  The angular velocity of the black hole is given by 
  $\Omega_\mathrm{h}\equiv a/(r_\mathrm{h}^2 + a^2)$.
  The ergosphere is characterized by the locus where the stationary
  Killing vector $(\partial_t)^\mu$ becomes null, i.e. $g_{tt}=0$, 
  and its radius is 
  $r_{\mathrm{ergo}}(\theta) \equiv M + \sqrt{M^2 - a^2\cos^2\theta}$.

\begin{figure}[t]
   \centering
   \includegraphics[width=7cm,clip]{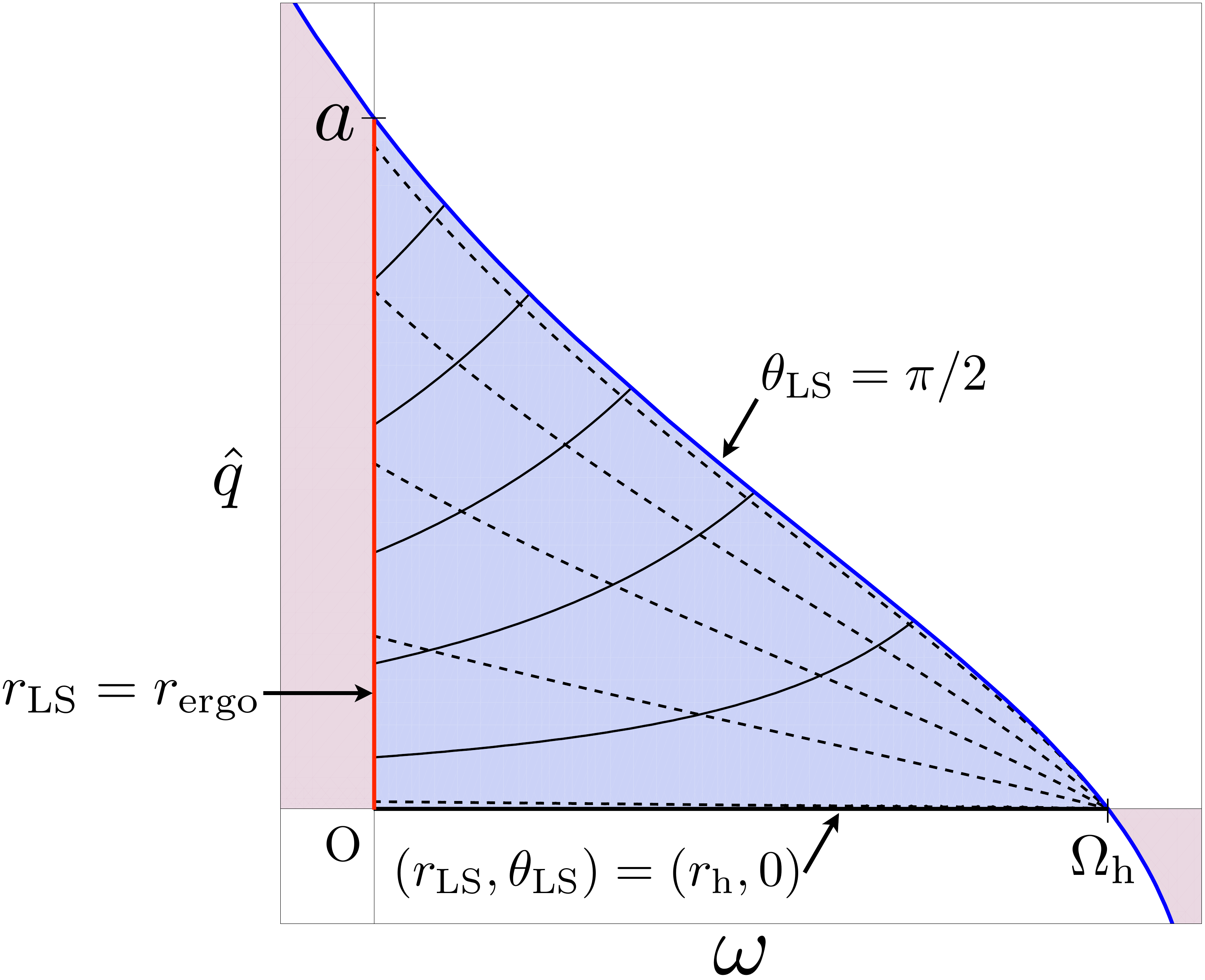}
   \caption{Typical shape of the allowed parameter region
   ($\omega,\hat{q}$) focusing on the energy-extraction region 
   (an example of $a/M=0.9$). 
   The shaded region is the allowed region, and in particular the
   triangular region 
   $\omega\hat{q}>0$ is where the energy extraction occurs. 
   The solid and dotted curves denote contour lines for the locus of the
   light surface $r_\mathrm{LS}$ and $\theta_\mathrm{LS}$, respectively.
   The radius of the light surface will coincide with that of the ergosphere 
   $r_\mathrm{LS}=r_\mathrm{ergo}$ on the vertical axis $\omega=0$ and
   that of the event horizon $r_\mathrm{LS}=r_\mathrm{h}$ on the
   horizontal axis $\hat{q}=0$.
   The curve represented by $\omega=\omega_\mathrm{eq}(\hat{q})$
   indicates that the light surface is located on the equatorial plane
   $\theta_\mathrm{LS}=\pi/2$; the horizontal axis that it 
   is on the rotation axis $\theta_\mathrm{LS} = 0$.
   }
   \label{fig:parameter_region}
  \end{figure}

  In this spacetime, the conditions (\ref{eq:electromagnetic_chi2}) and
  (\ref{eq:electromagnetic_q2}) at the light surface 
  ($r_\mathrm{LS}, \theta_\mathrm{LS}$) become 
  \begin{equation}
   \chi^2 = - \left[1 - \frac{2Mr}{\Sigma}(1-\omega a\sin^2\theta)^2
	      - \omega^2(r^2 + a^2)\sin^2\theta\right] = 0 ,
  \end{equation}
  and 
  \begin{equation}
   \hat{q}^2 = \Delta \sin^2\theta .
  \end{equation}
  Solving these equations in terms of $r$ and $\theta$ yields the locus
  of the light surface as%
  \footnote{
  As shown in Ref.~\cite{Kinoshita:2016lqd}, these equations have 
  another branch of solutions,  
  $r^-(\omega,\hat{q}) \equiv r_\mathrm{LS}(\omega,-\hat{q})$.
  This branch describes time-reversal processes rather than physically
  natural processes.
  }
  \begin{equation}
   r_\mathrm{LS}(\omega, \hat{q}) = \frac{M}{1+\hat{q}\omega}
    + \sqrt{\left(\frac{M}{1+\hat{q}\omega}\right)^2 - a(a-\hat{q})} ,
  \end{equation}
  and 
  \begin{equation}
   \sin\theta_\mathrm{LS} = \frac{|\hat{q}|}{\sqrt{\Delta(r_\mathrm{LS})}} .
  \end{equation}
  Here, we focus on the northern hemisphere and should take 
  $\pi - \theta_\mathrm{LS}$ in the southern hemisphere.
  In order to satisfy $\Delta(r_\mathrm{LS})>0$, namely, the
  light surface being located outside the horizon, and 
  $0 \le \sin^2\theta_\mathrm{LS} \le 1$, we obtain an allowed parameter
  region in terms of $(\omega, \hat{q})$.
  The allowed region can be described by the intervals in which $\omega$
  lies as 
  \begin{equation}
   \begin{aligned}
    \omega_\mathrm{axis} (\hat{q}) < \omega \le \omega_\mathrm{eq}(\hat{q}) \quad
    \text{for} \quad \hat{q}>0 ,\\
    \omega_\mathrm{eq}(\hat{q}) \le \omega < \omega_\mathrm{axis} (\hat{q}) \quad 
    \text{for} \quad \hat{q}<0 ,
   \end{aligned}
  \end{equation}
  where 
  \begin{equation}
   \omega_\mathrm{axis}(\hat{q}) \equiv - \frac{1}{\hat{q}}, \quad 
    \omega_\mathrm{eq} (\hat{q}) \equiv 
    \frac{a-\hat{q}}
    {2M^2 - \hat{q}(a-\hat{q}) + 2M\sqrt{M^2 - (a^2 - \hat{q}^2)}} .
  \end{equation}
  The boundaries represented by $\omega_\mathrm{eq}$ and
  $\omega_\mathrm{axis}$ indicate when the light surface is located at
  the equatorial plane ($\theta_\mathrm{LS} = \pi/2$) and when it approaches the
  rotation axis ($\theta_\mathrm{LS} \to 0$), respectively.
  Moreover, the radius of the light surface coincides with that of the
  ergosphere ($r_\mathrm{LS}=r_\mathrm{ergo}$) when $\omega=0$ 
  and that of the event horizon ($r_\mathrm{LS}=r_\mathrm{h}$) when
  $\hat{q}=0$ including $\omega=\Omega_\mathrm{h}$.
  The typical shape of the allowed region is shown in
  Fig.~\ref{fig:parameter_region}. 
  Since the region where the energy extraction occurs should be located
  in $\omega\hat{q}>0$, a necessary condition for the energy
  extraction is that a magnetic field line intersects the
  light surface inside the ergoregion of the Kerr black hole
  ($r_\mathrm{h}<r_\mathrm{LS}<r_\mathrm{ergo}$) and the 
  angular velocity of the magnetic field line is less than that of the
  black hole ($\omega<\Omega_\mathrm{h}$).
  It turns out that the curve of the upper boundary of the 
  energy-extraction region, represented by
  $\omega=\omega_\mathrm{eq}(\hat{q})$, should pass through
  $(\omega,\hat{q}) = (0,a), (\Omega_\mathrm{h},0)$. 
  Therefore, for an arbitrary $a$ the extraction rate of the specific
  energy can be bounded 
  by $\sim a\Omega_\mathrm{h}/4$, and such a maximum extraction rate will
  be achieved when the angular velocity of the magnetic field line is
  approximately 
  half of the black hole angular velocity near the equatorial plane
  ($\omega\simeq\Omega_\mathrm{h}/2$ and $\theta_\mathrm{LS}\simeq \pi/2$).
  It is worth noting that these extraction rates are irrelevant to the mass
  scale of the central black hole.

\begin{figure}[t]
   \begin{tabular}{cc}
    \begin{minipage}[b]{0.45\hsize}
     \centering
     \includegraphics[scale=1.0,clip]{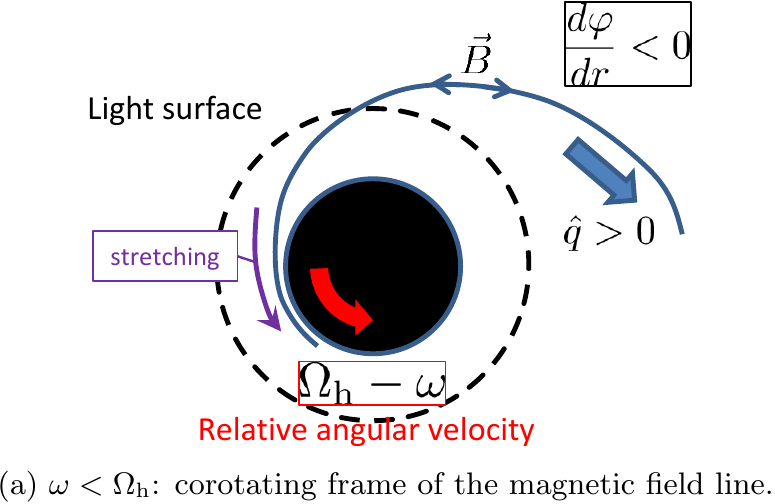}
    \end{minipage} &
    \begin{minipage}[b]{0.45\hsize}
     \centering
     \includegraphics[scale=1.0,clip]{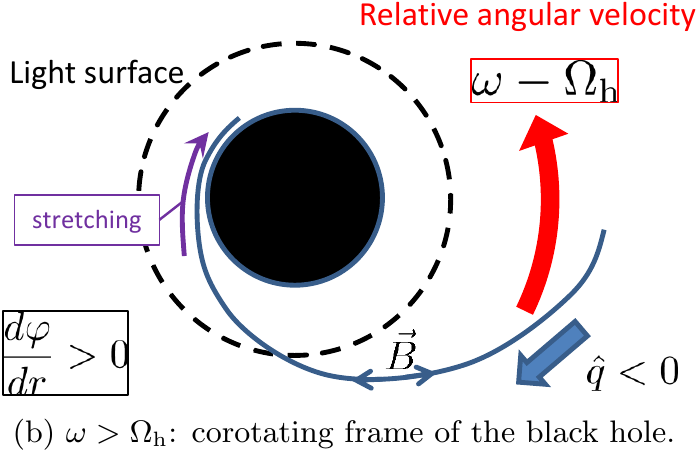}
    \end{minipage}
   \end{tabular}
   \caption{Relations among local configurations of the magnetic field
   line and directions of the specific angular-momentum flux $\hat{q}$.
   The left panel shows that the black hole is rotating
   faster than the magnetic field line ($\omega<\Omega_\mathrm{h}$) in the
   corotating frame of the field line;
   the right panel shows that the magnetic field line is rotating
   faster than the black hole ($\omega >\Omega_\mathrm{h}$) in the
   corotating frame of the black hole.
   The directions of the magnetic field are irrelevant.
   The dashed circle and the black disk depict, respectively, a light
   surface and a black hole viewed from the top along the rotation axis.
   The magnetic field line is stretching inside the light surface.
   }
   \label{fig:relation_q_and_fieldline}
  \end{figure}

  This necessary condition applies to each field sheet, i.e., each
  magnetic field line.
  It follows that each local configuration of the magnetic field line at
  the light surface governs whether energy and angular-momentum extraction
  can occur or not, i.e., the sign of $\omega\hat{q}$ and $\hat{q}$.
  Figure~\ref{fig:relation_q_and_fieldline} shows relations among local
  configurations of the magnetic field line and directions of the
  specific angular-momentum flux $\hat{q}$.
  From Eq.~(\ref{eq:electromagnetic_q_alt}) we find that the sign of
  $\hat{q}$ is directly connected with the sign of 
  $d\varphi/dr \equiv \mathcal{L}_\lambda \varphi/\mathcal{L}_\lambda r$
  regardless of the direction of $\lambda$.
  When the black hole is rotating faster than a magnetic field line, 
  $\omega(\psi)<\Omega_\mathrm{h}$, the magnetic field line is braking
  the rotation of the black hole, namely, extracting the angular momentum
  from the black hole, $\hat{q}>0$.
  When a magnetic field line is rotating faster than the black hole, 
  $\omega(\psi)>\Omega_\mathrm{h}$, the magnetic field line is
  accelerating the rotation of the black hole, namely, injecting the
  angular momentum into the black hole, $\hat{q}<0$.
  Furthermore, on the premise that the angular-momentum extraction has
  occurred $\hat{q}>0$, the magnetic field line can extract the
  rotational energy of the black hole $\omega(\psi)\hat{q}>0$
  if the light surface enters into the ergoregion
  $r_\mathrm{h} < r_\mathrm{LS} < r_\mathrm{ergo}$.

  As in the case of the rigidly rotating strings, this energy-extraction
  mechanism via the force-free electromagnetic fields can be simply
  interpreted as an analogy of the Penrose process.
  Outside the light surface, each field sheet is stationary with respect
  to each corotating vector 
  $\chi=\partial_t + \omega(\psi)\partial_\phi$, which is tangential to
  the field sheet.
  In other words, the configuration of the magnetic field line does not
  change in the corotating frame with angular velocity $\omega(\psi)$.
  However, inside the light surface the field sheet cannot be stationary
  because the corotating vector is spacelike.
  This means that the proper motion of each line element of the magnetic
  field line cannot follow the corotating angular velocity, and
  therefore the magnetic field line is stretching while its
  configuration remains unchanged.
  As a result, angular momentum associated with
  each line element is transferred toward the central black hole.
  When the black hole is rotating faster than the magnetic field line,
  this angular-momentum transfer will make the black hole spin down; 
  when the magnetic field line is rotating faster than the black hole, 
  it will make the black hole spin up.
  If angular-momentum transfer of the magnetic field line such that the
  black hole will be spun down occurs in the ergoregion, the magnetic field
  line can gain energy as a reaction to the angular-momentum transfer. 
  The energy gain will be transferred away from the black hole by the
  magnetic tension of the magnetic field line.
  This process is quite similar to the Penrose process for particles.
  Roughly speaking, the stretching part of the magnetic field lines plays
  the role of the infalling ``object'' in the Penrose process.
 
  The total amount of energy and angular-momentum flux is given by
  multiplying the specific quantities by an effective magnetic tension
  and integrating the contributions of every magnetic line.
  In fact, because $B\omega\hat{q}$ and $B\hat{q}$ are conserved on each
  field sheet, once values of $B$ on the light surface are given, we can
  obtain the total energy and angular-momentum flux by integrating 
  $B\omega\hat{q}$ and $B\hat{q}$ on the whole light surface as in
  Eq.~(\ref{eq:total_fluxes}). 
  To know the details of $B$, we have to solve the equations of motion and
  need global information on configurations of the magnetic field.
  However, the order of possible fluxes together with whether they are
  injecting or extracting has been locally determined by the
  specific quantities per unit tension, so that $B$ plays the role of a
  weight function of the magnetic tension.

  An average magnetic tension surrounding the black hole is defined by 
  \begin{equation}
   4\pi r_\mathrm{h}^2 \langle B^2\rangle \equiv 2\pi \int_{\mathrm{LS}} B|d\psi|.
  \end{equation}
  The energy-extraction rate can be estimated as 
  \begin{equation}
   \frac{d\mathcal{M}}{dt} = 
    \frac{\pi}{2} M^2 \langle B^2 \rangle u(\alpha)
    \int_{\mathrm{LS}} \frac{B|d\psi|}{2r_\mathrm{h}^2 \langle B^2\rangle}
    \frac{\omega \hat{q}}{\Omega_\mathrm{h} a/4},
  \end{equation}
  where $u(\alpha) \equiv \alpha^2 (1+\sqrt{1-\alpha^2})$ and 
  $\alpha \equiv a/M$.

 The whole of the Blandford--Znajek process can be classified into what is
 governed by local kinematics and what is governed by dynamics by
 solving the equations of motion.
 The former is the energy-extraction mechanism characterized by the
 relations between the specific energy and angular-momentum fluxes,
 $\omega\hat{q}$ and $\hat{q}$, and the locus of the light surface,
 which we have shown in this paper;
 the latter is global configurations of the
 magnetic field lines and the electric current, the value of the magnetic
 tension $B$, and so on.
 This fact suggests the following whole picture of the Blandford--Znajek
 process. 
 Depending on the environments in an outer region,
 the global configurations of
 the magnetic field lines associated with plasmas are dynamically
 determined according to the equations of motion.
 If only appropriate configurations of the magnetic field lines can be
 locally realized in an inner region around the ergoregion, the magnetic field
 lines start to extract the energy and angular momentum of the black
 hole by a kinematical mechanism irrelevant to situations in the outer
 region.
 For example, to obtain explicitly the directional dependence of the energy and
 angular-momentum fluxes in the far region, we have to solve
 globally the equations of motion for the electromagnetic fields and to
 know the entire configurations of the magnetic field lines.
 This implies that we should examine the details of the surrounding system, 
 i.e., the boundary conditions in the outer region, in order to explore the
 directional dependence in the far region where astrophysical jets are
 generated.
 However, in order to explain and understand the energy-extraction
 mechanism in the Blandford--Znajek process, we should consider just local
 kinematics in the inner region between the light surface and the
 ergosphere.
 The directional dependence of the specific energy and angular-momentum
 fluxes at the light surface has been shown in the contour of,
 e.g., Fig.~\ref{fig:parameter_region}.
 This provides the potential to generate the energy and angular-momentum
 fluxes in the Blandford--Znajek process. 
 We can see that the closer to the equatorial plane $\theta_\mathrm{LS}=\pi/2$ 
 the locus where the magnetic field lines intersect the light surface
 is, the larger specific energy flux $\omega\hat{q}$ the magnetic field
 lines can generate in general.

 \section{Discussion}
 
 In this paper we have shown that the essence of the energy extraction
 by the Blandford--Znajek process is local kinematics inside the
 ergoregion.
 Therefore, global configurations of the magnetic field line, such as
 whether the magnetic field lines can thread the event horizon or not, 
 should not be the essence of this process.
 The Znajek condition~\cite{Znajek1977}, which is well known as a condition to be
 satisfied at the event horizon, cannot be a necessary condition for the
 energy extraction. 
 This condition is a consistency condition for
 the force-free magnetic field lines to regularly cross the event horizon
 (an identical condition (\ref{eq:Znajek_string}) can be derived for the rigidly rotating strings).
 The main process for the energy extraction 
 occurs between the inner-light
 surface and the ergosphere, and, besides, the light surface is the
 causal boundary for this system.
 Therefore, 
 the energy extraction can occur even if
 this condition is not necessarily satisfied.
 For instance, it does not matter if the force-free condition for plasma
 has been violated at the event horizon.
 For the same reason, moreover, the so-called Meissner-like effect, in
 which 
 the extremal Kerr black hole tends to expel magnetic fields
 (see, e.g., Ref.~\cite{Komissarov:2007rc} and references therein),
 should not
 have a direct connection with the Blandford--Znajek process as long as
 we do not require an extra assumption at the horizon for a different
 reason.

 Throughout this paper we have discussed stationary and axisymmetric 
 force-free electromagnetic fields for the energy extraction.
 At the least, we need these assumptions between the inner-light surface and
 the ergosphere.
 If such assumptions are violated, it is expected that the energy
 extraction by the Blandford--Znajek process will become less efficient.
 The reason why the Blandford--Znajek process can efficiently extract the
 rotational energy from the black hole and produce a highly powerful
 energy flux is that a relativistic tension, which is equal to its
 energy density,
 carries the angular-momentum and energy flux along the magnetic field
 lines in the same way as Nambu--Goto strings.
 If the axisymmetry, stationariness, or force-freeness are violated, 
 the magnetic pressure or other matters begin to affect the
 angular-momentum and energy transfer and then alter the correspondence
 to the Nambu--Goto strings.
 This seems to be a disadvantage for the energy extraction.

 What is necessary for the Blandford--Znajek process to occur is toroidal
 magnetic fields winding the black hole at the light surface inside the
 ergoregion.
 As we showed in Eq.~(\ref{eq:electromagnetic_q_alt}), 
 the angular-momentum and energy fluxes depend on toroidal
 configurations of the magnetic field lines, namely, the toroidal component
 of the magnetic field.
 This is because toroidal magnetic fields take the central role in the
 angular-momentum transfer of the magnetic field lines.
 The poloidal components of the magnetic fields determine the directions
 of the angular-momentum and energy flows.

 For the Blandford--Znajek process, since magnetic field lines with 
 magnetic tension are essential and the transferred energy and angular
 momentum are purely electromagnetic, it seems that the electric current or
 plasma has only an 
 auxiliary role to sustain the magnetic fields.
 In fact, Ref.~\cite{Jacobson:2017xam} recently showed that 
 magnetic fields without plasma can extract the rotational
 energy of a black hole in lower spacetime dimensions.

 \begin{acknowledgments}
  We would like to thank Kenji~Toma and Yudai~Suwa 
  for useful comments.
  We would like to thank the participants of the workshop ``FuwakuBZ77''
  for helpful discussions. 
  This work was supported by JSPS KAKENHI Grant Numbers
  JP16K17704~(S.K.) and the MEXT-Supported Program for the Strategic
  Research Foundation at Private Universities, 2014-2017, S1411024~(T.I.). 
 \end{acknowledgments}

  \appendix

 \section{RIGIDLY ROTATING STRINGS}
 \label{app:string}
 
 In this appendix we summarize useful results for rigidly rotating
 strings around a rotating black hole, some of which were obtained in
 Ref.~\cite{Kinoshita:2016lqd}.

 In a stationary and axisymmetric spacetime (\ref{eq:metric}), 
 a stationary rotating string with angular velocity $\omega$ is embedded as 
 \begin{equation}
  t=\tau, \quad \phi = \omega \tau + \varphi(\sigma), \quad 
   r=r(\sigma), \quad \theta = \theta(\sigma) .
 \end{equation}
 The corotating (Killing) vector $\chi$ that is tangential to the world
 sheet becomes 
 \begin{equation}
  \chi = \frac{\partial}{\partial t} + \omega \frac{\partial}{\partial\phi}.
 \end{equation}
 The induced metric of the world sheet is given by 
 \begin{equation}
  h_{ab}d\sigma^a d\sigma^b
   = (g_{tt} + 2g_{t\phi}\omega + g_{\phi\phi}\omega^2)d\tau^2
   + 2\varphi'(g_{t\phi} + g_{\phi\phi}\omega) d\tau d\sigma
   + (g_{rr}r'^2 + g_{\theta\theta}\theta'^2 + g_{\phi\phi}\varphi'^2)
   d\sigma^2 ,
   \label{eq:string_hab}
 \end{equation}
 where the prime denotes a derivative with respect to $\sigma$.
 From this induced metric 
 the Lagrangian density for the Nambu--Goto string with unit tension is given by 
 \begin{equation}
  \mathscr{L}
   = - \sqrt{(g_{t\phi}^2 - g_{tt}g_{\phi\phi})\varphi'^2
   - \chi^2(g_{rr}r'^2 + g_{\theta\theta}\theta'^2)} ,
   \label{eq:string_Lagrangian}
 \end{equation}
 where $\chi^2 = \chi^\mu\chi^\nu g_{\mu\nu} = h_{\tau\tau}$.
 The light surface is characterized by 
 \begin{equation}
  \chi^2 = g_{tt} + 2g_{t\phi}\omega + g_{\phi\phi}\omega^2 = 0 ,
   \label{eq:string_chi2}
 \end{equation}
 and a consistency condition that the string can pass through the light
 surface regularly is given by 
 \begin{equation}
  q^2 = (g_{t\phi}^2 - g_{tt}g_{\phi\phi})|_{\chi^2 = 0} .
   \label{eq:string_q2}
 \end{equation}
 Here, $q$ is a conserved quantity characterizing the configuration of
 the rigidly rotating strings, which means the specific 
 angular-momentum flux per unit tension flowing on the string world sheet. 
 Its expression is written as 
 \begin{equation}
  q \equiv \frac{\partial \mathscr{L}}{\partial \varphi'}
   = - \frac{(g_{t\phi}^2 - g_{tt}g_{\phi\phi})\varphi'}
   {\sqrt{(g_{t\phi}^2 - g_{tt}g_{\phi\phi})\varphi'^2
   - \chi^2(g_{rr}r'^2 + g_{\theta\theta}\theta'^2)}} ,
   \label{eq:string_q}
 \end{equation}
 where the sign of $q$ has been defined so that $q>0$ describes a radially
 outward flux when $r'>0$.

 Using Eqs.~(\ref{eq:string_Lagrangian}) and (\ref{eq:string_q}), we
 have the following identity: 
 \begin{equation}
  q^2 \mathscr{L}^2 =
  (g_{t\phi}^2 - g_{tt}g_{\phi\phi}) [\mathscr{L}^2
    + \chi^2(g_{rr}r'^2 + g_{\theta\theta}\theta'^2)] .
    \label{eq:identity_q2L2}
 \end{equation}
 Suppose that 
 $(g_{t\phi}^2 - g_{tt}g_{\phi\phi})|_{r=r_\mathrm{h}} = 0$ and 
 $(g_{rr})^{-1}|_{r=r_\mathrm{h}} = 0$ should be satisfied 
 at the event horizon $r=r_\mathrm{h}$.
 The determinant of the metric components $g$ should not be degenerate,
 so that the equation 
 \begin{equation}
  (g_{t\phi}^2 - g_{tt}g_{\phi\phi})g_{rr} = -g/g_{\theta\theta} 
   \label{eq:det_g}
 \end{equation}
 is satisfied even at $r=r_\mathrm{h}$. 
 If the string can regularly pass through the event horizon, the volume
 element of the string world sheet, namely, the Lagrangian density 
 $\mathscr{L}$, should be finite and nonzero at the event horizon.
 Moreover, at $r=r_\mathrm{h}$ we find   
 \begin{equation}
  \chi^2 |_{r=r_\mathrm{h}} = g_{\phi\phi} 
   (\omega - \Omega_\mathrm{h})^2 ,
   \label{eq:chi2_horizon}
 \end{equation}
 where we have used $g_{t\phi}^2 = g_{tt}g_{\phi\phi}$ and 
 $\Omega_\mathrm{h} = - g_{t\phi}/g_{\phi\phi}$ at $r=r_\mathrm{h}$.
 Combining the above results (\ref{eq:identity_q2L2}), (\ref{eq:det_g}), 
 and (\ref{eq:chi2_horizon}), we obtain the following condition at the
 event horizon: 
 \begin{equation}
  q^2 = \left.(\omega - \Omega_\mathrm{h})^2 
	 \frac{(-g)}{\mathscr{L}^2}
	 \left(\frac{dr}{d\sigma}\right)^2
   \frac{g_{\phi\phi}}{g_{\theta\theta}}\right|_{r=r_\mathrm{h}} .
  \label{eq:Znajek_string}
 \end{equation}
 Note that, by definition, the combination 
 $\mathscr{L}d\sigma$ is invariant under
 reparameterization of the world-sheet coordinate $\sigma$.
 This condition is identical to the Znajek condition for force-free
 electromagnetic fields: 
 \begin{equation}
  I(\psi) = \left.(\omega - \Omega_\mathrm{h})\partial_\theta\psi 
   \sqrt{\frac{g_{\phi\phi}}{g_{\theta\theta}}}\right|_{r=r_\mathrm{h}} ,
  \label{eq:Znajek}
 \end{equation}
 which we can also obtain by evaluating Eq.~(\ref{eq:I2_B2_relation}) at the
 event horizon.


\begin{thebibliography}{99}


\bibitem{Blandford:1977ds} 
  R.~D.~Blandford and R.~L.~Znajek,
  ``Electromagnetic extractions of energy from Kerr black holes,''
  Mon.\ Not.\ Roy.\ Astron.\ Soc.\  {\bf 179}, 433 (1977).

\bibitem{MacDonald:1982zz} 
  D.~MacDonald and K.~S.~Thorne,
  ``Black-hole electrodynamics - an absolute-space/universal-time formulation,''
  Mon.\ Not.\ Roy.\ Astron.\ Soc.\  {\bf 198}, 345 (1982).

\bibitem{Komissarov:2008yh} 
  S.~S.~Komissarov,
  ``Blandford-Znajek mechanism versus Penrose process,''
  J.\ Korean Phys.\ Soc.\  {\bf 54}, 2503 (2009)
  [arXiv:0804.1912 [astro-ph]].

\bibitem{Koide:2014xpa} 
  S.~Koide and T.~Baba,
  ``Causal extraction of black hole rotational energy by various kinds of electromagnetic fields,''
  Astrophys.\ J.\  {\bf 792}, 88 (2014)
  [arXiv:1407.7088 [astro-ph.HE]].

\bibitem{Toma:2016jmz} 
  K.~Toma and F.~Takahara,
  ``Causal production of the electromagnetic energy flux and role of the negative energies in the Blandford-Znajek process,''
  PTEP {\bf 2016}, no. 6, 063E01 (2016)
  [arXiv:1605.03659 [astro-ph.HE]].


\bibitem{McKinney:2004ka} 
  J.~C.~McKinney and C.~F.~Gammie,
  ``A Measurement of the electromagnetic luminosity of a Kerr black hole,''
  Astrophys.\ J.\  {\bf 611}, 977 (2004)
  [astro-ph/0404512].

\bibitem{McKinney:2012vh} 
  J.~C.~McKinney, A.~Tchekhovskoy, and R.~D.~Blandford,
  ``General Relativistic Magnetohydrodynamic Simulations of Magnetically Choked Accretion Flows around Black Holes,''
  Mon.\ Not.\ Roy.\ Astron.\ Soc.\  {\bf 423}, 3083 (2012)
  [arXiv:1201.4163 [astro-ph.HE]].

\bibitem{Penna:2013rga} 
  R.~F.~Penna, R.~Narayan, and A.~Sadowski,
  ``General Relativistic Magnetohydrodynamic Simulations of Blandford-Znajek Jets and the Membrane Paradigm,''
  Mon.\ Not.\ Roy.\ Astron.\ Soc.\  {\bf 436}, 3741 (2013)
  [arXiv:1307.4752 [astro-ph.HE]].


\bibitem{Frolov:1996xw} 
  V.~P.~Frolov, S.~Hendy, and J.~P.~De Villiers,
  ``Rigidly rotating strings in stationary axisymmetric space-times,''
  Class.\ Quant.\ Grav.\  {\bf 14}, 1099 (1997)
  [hep-th/9612199].

\bibitem{Kinoshita:2016lqd} 
  S.~Kinoshita, T.~Igata, and K.~Tanabe,
  ``Energy extraction from Kerr black holes by rigidly rotating strings,''
  Phys.\ Rev.\ D {\bf 94}, no. 12, 124039 (2016)
  [arXiv:1610.08006 [gr-qc]].

\bibitem{Barrow:2017atq} 
  J.~D.~Barrow and G.~W.~Gibbons,
  ``Maximum magnetic moment to angular momentum conjecture,''
  Phys.\ Rev.\ D {\bf 95}, no. 6, 064040 (2017)
  [arXiv:1701.06343 [gr-qc]].

\bibitem{Dyson:1963}
  F.~Dyson, in {\em Interstellar Communication,}
  eds. A.~G.~Cameron
  (Benjamin, New York, 1963), Chap. 12.


\bibitem{Carter:1979} 
  B.~Carter, 
  in 
  {\em General Relativity: An Einstein Centenary Survey,}
  eds. S.~W.~Hawking and W.~Israel
  (Cambridge University Press, Cambridge, 1979),
  p.294-369.

\bibitem{Uchida:1997a}
  T.~Uchida,
  ``Theory of force-free electromagnetic fields. I. General theory,''
  Phys.\ Rev.\ E {\bf 56}, 2181 (1997).

\bibitem{Uchida:1997b}
  T.~Uchida,
  ``Theory of force-free electromagnetic fields. II. Configuration with symmetry,''
  Phys.\ Rev.\ E {\bf 56}, 2198 (1997). 

\bibitem{Gralla:2014yja} 
  S.~E.~Gralla and T.~Jacobson,
  ``Spacetime approach to force-free magnetospheres,''
  Mon.\ Not.\ Roy.\ Astron.\ Soc.\  {\bf 445}, no. 3, 2500 (2014)
  [arXiv:1401.6159 [astro-ph.HE]].

\bibitem{Christensson:1999tp} 
  M.~Christensson and M.~Hindmarsh,
  ``Magnetic fields in the early universe in the string approach to MHD,''
  Phys.\ Rev.\ D {\bf 60}, 063001 (1999)
  [astro-ph/9904358].

\bibitem{Semenov:2000vb} 
  V.~S.~Semenov,
  ``String mechanism for energy extraction from a Kerr black hole,''
  Phys.\ Scripta {\bf 62}, 123 (2000).

\bibitem{Semenov:2004ib} 
  V.~Semenov, S.~Dyadechkin and B.~Punsly,
  ``Simulations of jets driven by black hole rotation,''
  Science {\bf 305}, 978 (2004)
  [astro-ph/0408371].

\bibitem{Frolov:1998wf} 
  V.~P.~Frolov and I.~D.~Novikov,
  {\em Black hole physics: Basic concepts and new developments,}
  (Fundamental theories of physics. 96). 

\bibitem{Takahashi:1990bv} 
  M.~Takahashi, S.~Nitta, Y.~Tatematsu, and A.~Tomimatsu,
  ``Magnetohydrodynamic flows in Kerr geometry: Energy extraction from black holes,''
  Astrophys.\ J.\  {\bf 363}, 206 (1990).

\bibitem{Contopoulos:2012py} 
  I.~Contopoulos, D.~Kazanas, and D.~B.~Papadopoulos,
  ``The Force-Free Magnetosphere of a Rotating Black Hole,''
  Astrophys.\ J.\  {\bf 765}, 113 (2013)
  [arXiv:1212.0320 [astro-ph.HE]].

\bibitem{Thoelecke:2017jrz} 
  K.~Thoelecke, M.~Takahashi, and S.~Tsuruta,
  ``Effects of inner Alfv\'en surface location on black hole energy extraction in the limit of a force-free magnetosphere,''
  Phys.\ Rev.\ D {\bf 95}, no. 6, 063008 (2017)
  [arXiv:1702.07322 [astro-ph.HE]].

\bibitem{Komissarov:2002dj} 
  S.~S.~Komissarov,
  ``On the nature of the Blandford-Znajek mechanism,''
  astro-ph/0211141.

\bibitem{Ruiz:2012te} 
  M.~Ruiz, C.~Palenzuela, F.~Galeazzi, and C.~Bona,
  ``The Role of the ergosphere in the Blandford-Znajek process,''
  Mon.\ Not.\ Roy.\ Astron.\ Soc.\  {\bf 423}, 1300 (2012)
  [arXiv:1203.4125 [gr-qc]].

\bibitem{Lasota:2013kia} 
  J.-P.~Lasota, E.~Gourgoulhon, M.~Abramowicz, A.~Tchekhovskoy, and R.~Narayan,
  ``Extracting black-hole rotational energy: The generalized Penrose process,''
  Phys.\ Rev.\ D {\bf 89}, no. 2, 024041 (2014)
  [arXiv:1310.7499 [gr-qc]].

\bibitem{Compere:2016xwa} 
  G.~Comp\`ere, S.~E.~Gralla, and A.~Lupsasca,
  ``Force-Free Foliations,''
  Phys.\ Rev.\ D {\bf 94}, no. 12, 124012 (2016)
  [arXiv:1606.06727 [math-ph]].

\bibitem{Schild:1976vq} 
  A.~Schild,
  ``Classical Null Strings,''
  Phys.\ Rev.\ D {\bf 16}, 1722 (1977).

\bibitem{Kastrup:1979fz} 
  H.~A.~Kastrup,
  ``Relativistic Strings and Electromagnetic Flux Tubes,''
  Phys.\ Lett.\  {\bf 82B}, 237 (1979).

\bibitem{Rinke:1979mw} 
  M.~Rinke,
  ``The Relation Between Relativistic Strings and Maxwell Fields of Rank 2,''
  Commun.\ Math.\ Phys.\  {\bf 73}, 265 (1980).

\bibitem{Nambu:1981gt} 
  Y.~Nambu,
  ``Effective Abelian Gauge Fields,''
  Phys.\ Lett.\  {\bf 102B}, 149 (1981).

\bibitem{Carter:2000wv} 
  B.~Carter,
  ``Essentials of classical brane dynamics,''
  Int.\ J.\ Theor.\ Phys.\  {\bf 40}, 2099 (2001)
  [gr-qc/0012036].

\bibitem{Znajek1977} 
  R.~L.~Znajek,
  ``Black hole electrodynamics and the Carter tetrad,''
  Mon.\ Not.\ Roy.\ Astron.\ Soc.\  {\bf 179}, 457 (1977).

\bibitem{Komissarov:2007rc} 
  S.~S.~Komissarov and J.~C.~McKinney,
  ``Meissner effect and Blandford-Znajek mechanism in conductive black hole magnetospheres,''
  Mon.\ Not.\ Roy.\ Astron.\ Soc.\  {\bf 377}, L49 (2007)
  [astro-ph/0702269].

\bibitem{Jacobson:2017xam} 
  T.~Jacobson and M.~J.~Rodriguez,
  ``Blandford-Znajek process in vacuo and its holographic dual,''
  arXiv:1709.10090 [hep-th].

 \end{thebibliography}
\end{document}